\input harvmac


\input epsf

\newcount\figno
\figno=0
\def\fig#1#2#3{
\par\begingroup\parindent=0pt\leftskip=1cm\rightskip=1cm\parindent=0pt
\baselineskip=11pt
\global\advance\figno by 1
\midinsert
\epsfxsize=#3
\centerline{\epsfbox{#2}}
\vskip 12pt
{\bf Figure \the\figno:} #1\par
\endinsert\endgroup\par
}
\def\figlabel#1{\xdef#1{\the\figno}}


\def\doublefig#1#2#3#4#5{
\par\begingroup\parindent=0pt\leftskip=1cm\rightskip=1cm\parindent=0pt
\baselineskip=11pt
\global\advance\figno by 1
\midinsert
\epsfxsize=#4
\centerline{\epsfbox{#2}\hskip0.2in\epsfxsize=#5\epsfbox{#3}}
\vskip 12pt
{\bf Figure \the\figno:} #1\par
\endinsert\endgroup\par
}
\def\figlabel#1{\xdef#1{\the\figno}}


\noblackbox

\def\IZ{\relax\ifmmode\mathchoice
{\hbox{\cmss Z\kern-.4em Z}}{\hbox{\cmss Z\kern-.4em Z}}
{\lower.9pt\hbox{\cmsss Z\kern-.4em Z}} {\lower1.2pt\hbox{\cmsss
Z\kern-.4em Z}}\else{\cmss Z\kern-.4em Z}\fi}

\font\cmss=cmss10 \font\cmsss=cmss10 at 7pt
\def\IR{\relax{\rm I\kern-.18em R}}
\def\inbar{\,\vrule height1.5ex width.4pt depth0pt}
\def\IC{\relax\hbox{$\inbar\kern-.3em{\rm C}$}}
\def\IR{\relax{\rm I\kern-.18em R}}
\def\IP{\relax{\rm I\kern-.18em P}}
\def\IZ{\relax{\rm Z\kern-.34em Z}}
\def\One{{1\hskip -3pt {\rm l}}}

\def\frac#1#2{{#1 \over #2}}


\def\journal#1&#2(#3){\unskip, \sl #1\ \bf #2 \rm(19#3) }
\def\andjournal#1&#2(#3){\sl #1~\bf #2 \rm (19#3) }

\def\d{\partial}

%

%
\catcode`\@=11
\def\slash#1{\mathord{\mathpalette\c@ncel{#1}}}
\overfullrule=0pt

\def\NN{{\cal N}}

\def\ZZ{{\cal Z}}

\def\underrel#1\over#2{\mathrel{\mathop{\kern\z@#1}\limits_{#2}}}

\catcode`\@=12

\def\({\left(}
\def\){\right)}
\def\[{\left[}
\def\]{\right]}
\def\<{\langle}
\def\>{\rangle}
\def\half{{1\over 2}}
\def\d{\partial}
\def\tr{{\rm Tr}}
\def\|{\biggl|}

\def\det{{\rm det}}
\def\exp{{\rm exp}}

\def\bk{{\bf k}}
\def\bp{{\bf p}}
\def\bx{{\bf x}}

\def\bA{{\bf A}}
\def\ba{{\bf a}}
\def\by{{\bf y}}

\def\ooint{\relax{\int\kern-.9em ^{\rm O}\kern-.75em _{\rm O}}}
\def\uoint{\relax{\int\kern-.87em ^{\rm O}}}
\def\doint{\relax{\int\kern-.97em _{\rm O}}}
\def\square{\kern1pt\vbox{\hrule height 1.2pt\hbox{\vrule width 1.2pt\hskip 3pt \vbox{\vskip 6pt}\hskip 3pt\vrule width 0.6pt}\hrule height 0.6pt}\kern1pt}

\def\li2{{\rm Li_2}}


%


\lref\MaldacenaMH{
  J.~M.~Maldacena and J.~G.~Russo,
  ``Large N limit of non-commutative gauge theories,''
  JHEP {\bf 9909}, 025 (1999)
  [arXiv:hep-th/9908134].
}

\lref\HashimotoUT{
  A.~Hashimoto and N.~Itzhaki,
  ``Non-commutative Yang-Mills and the AdS/CFT correspondence,''
  Phys.\ Lett.\  B {\bf 465}, 142 (1999)
  [arXiv:hep-th/9907166].
}

\lref\BerkovitsIC{
  N.~Berkovits and J.~Maldacena,
  ``Fermionic T-Duality, Dual Superconformal Symmetry, and the Amplitude/Wilson
  Loop Connection,''
  JHEP {\bf 0809}, 062 (2008)
  [arXiv:0807.3196 [hep-th]]; N.~Beisert, R.~Ricci, A.~A.~Tseytlin and M.~Wolf,
  ``Dual Superconformal Symmetry from AdS5 x S5 Superstring Integrability,''
  Phys.\ Rev.\  D {\bf 78}, 126004 (2008)
  [arXiv:0807.3228 [hep-th]].
}

\lref\AldayHR{
  L.~F.~Alday and J.~M.~Maldacena,
  ``Gluon scattering amplitudes at strong coupling,''
  JHEP {\bf 0706}, 064 (2007)
  [arXiv:0705.0303 [hep-th]].
}

\lref\GrossBA{
  D.~J.~Gross, A.~Hashimoto and N.~Itzhaki,
  ``Observables of non-commutative gauge theories,''
  Adv.\ Theor.\ Math.\ Phys.\  {\bf 4}, 893 (2000)
  [arXiv:hep-th/0008075].
}

\lref\SeibergVS{
  N.~Seiberg and E.~Witten,
  ``String theory and noncommutative geometry,''
  JHEP {\bf 9909}, 032 (1999)
  [arXiv:hep-th/9908142].
}

\lref\McGreevyKT{
  J.~McGreevy and A.~Sever,
  ``Quark scattering amplitudes at strong coupling,''
  JHEP {\bf 0802}, 015 (2008)
  [arXiv:0710.0393 [hep-th]].
}

\lref\McGreevyZY{
  J.~McGreevy and A.~Sever,
  ``Planar scattering amplitudes from Wilson loops,''
  JHEP {\bf 0808}, 078 (2008)
  [arXiv:0806.0668 [hep-th]].
}

\lref\PolyakovJU{
  A.~M.~Polyakov,
  ``The wall of the cave,''
  Int.\ J.\ Mod.\ Phys.\  A {\bf 14}, 645 (1999)
  [arXiv:hep-th/9809057].
}

\lref\WittenWY{
  E.~Witten,
  ``AdS/CFT correspondence and topological field theory,''
  JHEP {\bf 9812}, 012 (1998)
  [arXiv:hep-th/9812012].
}

\lref\tHooftSZ{
  G.~'t Hooft,
  ``Some Twisted Selfdual Solutions For The Yang-Mills Equations On A
  Hypertorus,''
  Commun.\ Math.\ Phys.\  {\bf 81}, 267 (1981).
}

\lref\AlishahihaCI{
  M.~Alishahiha, Y.~Oz and M.~M.~Sheikh-Jabbari,
  ``Supergravity and Large N Noncommutative Field Theories,''
  JHEP {\bf 9911}, 007 (1999)
  [arXiv:hep-th/9909215].
}

\lref\MaldacenaIM{
  J.~M.~Maldacena,
  ``Wilson loops in large N field theories,''
  Phys.\ Rev.\ Lett.\  {\bf 80}, 4859 (1998)
  [arXiv:hep-th/9803002].
}

\lref\GrossGE{
  D.~J.~Gross and J.~L.~Manes,
  ``THE HIGH-ENERGY BEHAVIOR OF OPEN STRING SCATTERING,''
  Nucl.\ Phys.\  B {\bf 326}, 73 (1989).
}

\lref\StrasslerZR{
  M.~J.~Strassler,
  ``Field theory without Feynman diagrams: One loop effective actions,''
  Nucl.\ Phys.\  B {\bf 385}, 145 (1992)
  [arXiv:hep-ph/9205205].
}

\lref\LLL{
J.~S.~Schwinger,
 ``On gauge invariance and vacuum polarization,"
Phys.\ Rev.\  {\bf 82}, 664 (1951);
E.~V.~Gorbar and V.~A.~Miransky,
``Relativistic field theories in a magnetic background as noncommutative field theories,"
Phys.\ Rev.\ D {\bf 70}, 105007 (2004) [arXiv:hep-th/0407219].
}

\lref\SalimKY{
  A.~J.~Salim and N.~Sadooghi,
  ``Dynamics of O(N) model in a strong magnetic background field as a  modified
  noncommutative field theory,''
  Phys.\ Rev.\  D {\bf 73}, 065023 (2006)
  [arXiv:hep-th/0602023].
}

\lref\PolyakovCA{
  A.~M.~Polyakov,
  ``Gauge Fields As Rings Of Glue,''
  Nucl.\ Phys.\  B {\bf 164}, 171 (1980).
}

\lref\AldayHE{
  L.~F.~Alday and J.~Maldacena,
  ``Comments on gluon scattering amplitudes via AdS/CFT,''
  JHEP {\bf 0711}, 068 (2007)
  [arXiv:0710.1060 [hep-th]].
}

\lref\ConnesCR{
  A.~Connes, M.~R.~Douglas and A.~S.~Schwarz,
  ``Noncommutative geometry and matrix theory: Compactification on tori,''
  JHEP {\bf 9802}, 003 (1998)
  [arXiv:hep-th/9711162]; M.~R.~Douglas and C.~M.~Hull,
 ``D-branes and the noncommutative torus,''
  JHEP {\bf 9802}, 008 (1998)
  [arXiv:hep-th/9711165]; F.~Ardalan, H.~Arfaei and M.~M.~Sheikh-Jabbari,
  ``Noncommutative geometry from strings and branes,''
  JHEP {\bf 9902}, 016 (1999)
  [arXiv:hep-th/9810072]; M.~M.~Sheikh-Jabbari,
  ``Super Yang-Mills theory on noncommutative torus from open strings
  interactions,''
  Phys.\ Lett.\  B {\bf 450}, 119 (1999)
  [arXiv:hep-th/9810179].
}

\lref\Privetconversation{
 J.~Maldacena, private discussion.
}

\lref\BerkovitsZK{
  N.~Berkovits,
  ``ICTP lectures on covariant quantization of the superstring,''
  arXiv:hep-th/0209059.
}

\lref\FilkDM{
  T.~Filk,
  ``Divergencies in a field theory on quantum space,''
  Phys.\ Lett.\  B {\bf 376}, 53 (1996).
}

\lref\BrinkNB{
  L.~Brink and J.~H.~Schwarz,
  ``Quantum Superspace,''
  Phys.\ Lett.\  B {\bf 100}, 310 (1981).
}

\lref\Inprogress{
A.~Sever, in progress.
}

\Title{\vbox{\baselineskip12pt}} {\vbox{ \centerline{Non-commutative holography and} \centerline{scattering amplitudes in a large magnetic background} \smallskip
\smallskip
\smallskip
}} \centerline{Amit Sever}
\bigskip
\centerline{Perimeter Institute for Theoretical Physics}\centerline{Waterloo,
Ontario N2J 2W9, Canada} 

\bigskip
\bigskip
\bigskip
\noindent

We study planar gluon scattering amplitudes and Wilson loops in non-commutative gauge theory. Our main results are:
\item{1.} We find the map between observables in non-commutative gauge theory and their holographic dual. In that map, the region near the boundary of the gravitational dual describes the physics in terms of T-dual variables. 
\item{2.} We show that in the presence of a large magnetic background and a UV regulator, a planar gluon scattering amplitude reduces to a complex polygon Wilson loop expectation value, dressed by a tractable polarization dependent factor.

\Date{January 2009}

\newsec{Introduction and summary}

Gauge theories on non-commutative spaces (NCYM) can arise in certain limits of string theory \refs{\ConnesCR, \SeibergVS}. These have gravity duals whose near horizon region describes the large $N$ limit of the NCYM theories \refs{\HashimotoUT,\MaldacenaMH}. So far, we had only a poor understanding of how to do holography in these backgrounds. The main problem is that the string metric shrinks near the boundary as apposed to the lore we are used to from the AdS/CFT duality, where it expands. 

In this paper, we study planar gluon scattering amplitudes and Wilson loops in these holographic dual descriptions.\foot{A relation between the {\it scattering amplitude} $\leftrightarrow$ {\it Wilson loop} duality and NCYM was suggested in \AldayHR\ (see footnote 7) and in \Privetconversation.} We find that the region near the boundary of the gravitational dual describes the physics in term of T-dual variables.\foot{The precise meaning of ``T-dual variables" will be made clear in sections 3 and 5.} In that sense, turning on non-commutativity can be thought of as a scale dependent rotation between momentum and position (loop) space. Such rotation preserves the number of observables.

This study leads us to consider planar gluon scattering amplitudes with a UV regulator in a large magnetic field that couples to the boundary of planar diagrams. We show that in the limit where the magnetic backgroung is scaled to infinity, computing such scattering amplitude reduces to the computation of a complex polygon Wilson loop expectation value, dressed by a tractable polarization dependent factor. The difference between the real polygon loop made of the gluon momenta and the complex polygon loop is that each edge of the real polygon is replaced by two complex null edges. 

The paper is organized as follows: In section 2 we set our conventions and review the definitions of planar gluon scattering amplitudes and Wilson loops in NCYM. Up to an overall phase, these are blind to the non-commutativity scale. In section 3 we describe their gravitational duals. Agreement with the field theory results is found. The form of the gravity dual leads us in section 4, to go back to field theory and we study planar gluon scattering amplitude in a large magnetic background. In section 5 we briefly discuss closed string observables in backgrounds with NCYM dual.

\newsec{Planar scattering amplitudes and Wilson loops in Non-commutative gauge theories}

What are the definitions of planar scattering amplitudes and Wilson loops in NCYM? There is no unique answer to that question, as any observable that in the commutative limit reduces to the Wilson loop or the amplitude may serve as a valid definition. Here, we take the conservative approach in which all definitions remain the same as in ordinary gauge theory (OYM), apart form the multiplication which is replaced by the Moyal $\star$-product and the $\star$-ordering is kept the same as the color ordering. In the planar limit, apart from an overall phase factor, the perturbative result will be independent of the non-commutativity scale $\theta$ \FilkDM. In the next section, we will match these to observables in the gravity dual and confirm the perturbative result at strong coupling.

\subsec{Setup}

Here we consider NCYM that arise in the near horizon limit of $N$ D3-branes with constant NS $B$-field and $(2,2)$ signature.\foot{Note that the name $D3$ brane is a bit inappropriate here as we work in $(2,2)$ signature.} We turn on the same amount of non-commutativity in the time-time and space-space directions. We parametrize the non-commutativity as
\eqn\noncommutativity{[x^\mu,x^\nu]=i\theta^{\mu\nu}=i\theta M^{\mu\nu}~,}
where $\theta$ is the non-commutative scale and
\eqn\Mmatrix{M=\(\matrix{0&1&&\cr -1&0&&\cr && 0&1\cr &&-1&0}\)~.}
Note that $M$ maps null vectors into null vectors . We will use the following notation for the contraction with the $\theta$ tensor
$$\theta^{\mu\nu}k_\mu p_\nu=\theta M^{\mu\nu}k_\mu p_\nu=\bk\wedge\bp~,$$
where bold face letters stands for four vectors.

In addition, whenever discussing the gluon scattering amplitude, we will consider only the case of spacelike momentum transfer in all channels. The term ``dual" refers to the holographic dual (as appose to the T-dual).

\subsec{Wilson loops in NCYM}

Let 
\eqn\Wilson{\<W\[\bx(\cdot)\]\>_{\mu_{UV}}={1\over N}\<\int \tr P e^{i\int ds \bA(\bx(s))\cdot\dot\bx(s)}\>_{\mu_{UV}}}
be a Wilson loop in $U(N)$ gauge theory along a closed contour parametrized by $\bx(\cdot)$ and $\mu_{UV}$ a UV cutoff. We defined the non-commutative generalization of \Wilson\ to be
\eqn\conservativeWilson{\< W_\star\[\bx(\cdot)\]\>_{\mu_{UV}}={1\over N\,\rm Vol}\<\int d^4x_0\,\tr P_\star e^{i\int ds \bA(\bx(s)+\bx_0)\cdot\dot\bx(s)}\>_{\mu_{UV}}~,}
where ${\rm Vol}$ stands for the space-time volume and the $P_\star$ stands for star path ordering. At $\theta=0$ the integration over $\bx_0$ just gives an overall volume factor that cancels against $1/{\rm Vol}$ and \conservativeWilson\ reduces to \Wilson. For $\theta\ne 0$, only the integrated Wilson loop is gauge invariant. 

In the planar limit 
\eqn\Wilsoninvariant{\< W_\star\[\bx(\cdot)\]\>_{\mu_{UV}}=\< W\[\bx(\cdot)\]\>_{\mu_{UV}}}
is independent of the non-commutativity scale $\theta$. More generally, the planar expectation value of any integrated single trace observable made of operators that are color ordered and starred ordered in the same way is independent of the non-commutativity parameter. The reason is that at any order in perturbation theory, one is summing over correlation functions of say, $m$ gauge field operators starred and color contracted in the same order. In momentum space, this is the same as a vacuum to vacuum planar diagram with a $m$-vertex. Planar vacuum to vacuum diagrams are independent of the non-commutativity parameter \FilkDM.

\subsec{Gluon scattering amplitudes in NCYM}

Let $\varphi(\bx)$ be some massless field in the adjoint representation. The partial scattering amplitude of states created by $\varphi $ with momentum $\bk_{i=1,\dots,n}$ is defined through the LSZ formula to be
$$A(\bk_1,\dots,\bk_n)=\bk_1^2\,\bk_2^2\dots\bk_n^2\,\<\tr[\tilde \varphi(\bk_1)\tilde \varphi(\bk_2)\dots\tilde \varphi(\bk_n)]\>_{\mu_{IR}} $$
where $\bk_i^2$ stands for the inverse propagator, the external momentum are taken on-shell $\bk_i^2=0$, $\mu_{IR}$ is an IR cutoff and we have suppressed all other quantum numbers of the asymptotic states (like polarization and global charges).\foot{If $\varphi $ is a gluon then the exact form of the inverse propagator will depend on the gauge chosen.}

In the non-commutative theory we would like to scatter states created by the operator
\eqn\concervedop{\varphi(\bx)\star e^{i\bk\cdot\bx}~.}
Apart from carrying momentum, the operator $e^{i\bk\cdot\hat\bx}$ generates a translation by $\theta\bk$. Under gauge transformation \concervedop\ transforms as:
$$\varphi(\bx)\star e^{i\bk\cdot\bx}\quad\to\quad U(\bx)\star \varphi(\bx)\star e^{i\bk\cdot\bx}\star U^\dagger(\bx+\theta\bk)~.$$
That is how an extended object like a Wilson line transforms and is what one may expect from the stringy picture for an open string carrying ``winding". In that sense, the operator \concervedop\ is carrying both, momentum and ``winding". After integrating over $\bx$, we see that we can drop the star in \concervedop
$$\tilde \varphi_\star(\bk)=\int d^4x\, \varphi(\bx)\star e^{i\bk\cdot\bx}=\int d^4x\, \varphi(\bx) e^{i\bk\cdot\bx}=\tilde \varphi(\bk)~,$$
so we are scattering the ordinary momentum modes. That is, we defined the non-commutative scattering amplitude as:
\eqn\convervedamp{A_\star\equiv\bk_1^2\,\bk_2^2\dots\bk_n^2\,\<\tr[\tilde \varphi(\bk_1)\star\tilde \varphi(\bk_2)\star\dots\star\tilde \varphi(\bk_n)]\>_\star~.}
In the planar limit, apart from an overall phase factor, \convervedamp\ is independent of the non-commutativity parameter $\theta$ \FilkDM. That is
\eqn\planarrelation{A_\star=e^{i\Phi}A~,}
where the phase $\Phi$ is
\eqn\phase{\Phi=\half\sum_{i<j}\bk_i\wedge\bk_j~.}

We conclude that as far as the planar Wilson loop and scattering amplitude are considered, turning on non-commutativity is boring. That is because apart from an overall phase for the scattering amplitude, nothing is changing. Nevertheless, things will become interesting where, in section 4, we will think about momentum space as describing a gauge theory in position space.

\subsec{Scattering amplitude from Wilson loops in planar NCYM}

In \McGreevyZY\ an expression for the QCD planar gluon scattering amplitude in terms of Wilson loops was given. Here, for simplicity, we concentrate on a scalar contribution to 1PI planar graphs \PolyakovJU:
\eqn\ScattWilson{A_n^{\rm 1PI}=\int{dT\over
T}\NN\int[D\bx(\cdot)]_1e^{-\int_0^Tds\(\half\dot\bx^2+m^2\)}\prod_i\int_{s_{i-1}}^T ds_i\, \varepsilon_i\cdot\dot\bx(s_i)e^{ i\bk_i \cdot \bx(s_i)}\<W[\bx(\cdot)]\>~,}
where $m$ is a worldline mass, introduced as an IR regulator, $\int[D\bx]_1$ is an integral over all closed curves defined with respect to a trivial worldline metric. The normalization constant $\NN$ is defined such that in the continuum limit
$$\NN\int[D\bx(\cdot)]_1e^{-\half\int_0^T\dot\bx^2 ds}=[2\pi
T]^{-D/2}$$
 and $D=4$ is the number of spacetime dimensions. As in open string theory, the terms
$$V_g(\varepsilon_i,\bk_i;s_i)=\varepsilon_i^\mu{dx_\mu\over ds}e^{i \bk_i \cdot \bx(s_i)}$$
are worldline gluon vertex insertions ordered and integrated along the loop. Each closed loop is weighted by the expectation value of the corresponding (planar) Wilson loop $\<W[x(\cdot)]\>$. That equation \ScattWilson\ has a straight-forward generalization for the scattering of adjoint matter fields as well as to the spinor and vector worldline theories. Everything we say below also holds for these generalizations.

%
%

The momentum dependent phases in \ScattWilson\ can be written as:
\eqn\Polygonloop{\eqalign{i\sum_i\bk_i\cdot\bx(s_i)=&i\int ds \sum_i\bk_i\cdot\bx(s)\delta(s-s_i)=-i\int ds \sum_i\bk_i\cdot\dot\bx(s)\theta(s-s_i)\cr\equiv& -i\int ds \bp(s)\cdot\dot\bx(s)=-i\oint\bp\cdot d\bx~.}}
Using \Polygonloop, the planar scattering amplitude \ScattWilson\ was interpreted as a Fourier transform in loop space \PolyakovJU\ to the momentum polygon loop
\eqn\polygon{\bp(s)=\sum_i\bk_i\theta(s-s_i)~.}

One can repeat the derivation of \ScattWilson\ in NCYM. Again, the result is the same as in the ordinary YM theory apart from the overall phase \phase. It can be written as
\eqn\momentummagnetic{\Phi=\half\sum_{i<j}k_i^\mu\theta_{\mu\nu}k_j^\nu=\half\int ds\, \theta_{\mu\nu}p^\mu\dot p^\nu~.}
Therefore, turning on non-commutativity can be thought of as turning on a ``background magnetic field" in momentum space. Things will become more interesting when we will think of the momentum space as (T-dual) position space \AldayHR. Before doing so, we first describe the above observables in the dual string theory picture.

\newsec{The dual string theory picture}

\subsec{Gravity dual of NCYM}

Turning on non-commutativity changes the gravitational dual background. It arises from the near horizon limit of D3-branes with a constant NS $B$ field in the same limit as the field theory one \SeibergVS. Here, we turn on $B$-field along all the D3 directions (time-time and space-space in the $(2,2)$ signature). The gravity dual description of NCYM is \refs{\HashimotoUT,\MaldacenaMH}:
\eqn\background{{\rm NS:\qquad}\eqalign{ds^2=&\alpha'\sqrt\lambda\[u^2h\(-dx_{-1}^2-dx_0^2+dx_1^2+dx_2^2\)+{du^2\over u^2}+d\Omega_5^2\]~,\cr  e^{\phi}=&gh~,\qquad\alpha'B_{-1\,0}=\alpha'B_{1\,2}=\alpha'\lambda\theta u^4h~,\qquad h={1\over1+\lambda\theta^2u^4}~,}}
\smallskip \smallskip
\eqn\RRbackground{{\rm RR:\qquad}\chi=i{\lambda\theta^2\over g}u^4~,\quad A_{-1\,0}=A_{1\,2}=i{B_{12}\over g}~,\quad F_{-1012u}=i{\alpha'^2\lambda h^2\over g}\d_uu^4 ~,}
where $\phi,B$ are the NS fields, $A,F,\chi$ are the RR fields and $\theta$ is the non-comutativity scale of the dual NCYM \noncommutativity. This solution reduces to the $AdS_5\times S^5$ solution for small $u$, which correspond to the IR regime of the gauge theory. That is an intuitive picture where NCYM reduces to OYM theory at long distances and the non-commutativity becomes more and more apparent at short distances. Near the boundary ($u\to\infty$) the string metric looks again as AdS metric near the Poincare horizon, the string coupling runs to zero and the $B$-field approaches a constant value
$$B_\infty={\alpha'\over\theta}~.$$
We will call the background \background\ NCAdS. 

For simplicity, from now on, we set $\alpha'=R_{AdS}=1$ (so $\lambda=1$ as well). We also drop the RR fields and the sphere part of the metric since these will not play an important role.

Few comments are in order:
\item{-} If the theory is compactified on a torus of size $L$, then we have a dimensionless parameter $\theta/L^2$ measuring the amount of non-commutativity. In the non-compact case we are interested in here, there is no such dimensionless parameter and by rescaling $u$ and $\bx$ we can change the dimensionful value of $\theta$ as we like, keeping the background \background, \RRbackground\ the same. So before introducing another scale into the problem (like a cutoff), there is no meaningful sense in which we can say that $\theta$ is small or large.

\item{-} The NCAdS background \background\ is invariant under 
\eqn\shifted{u\to{1\over \theta u}~,\qquad B\to B_\infty-B~,\qquad e^\phi\to g-e^\phi~.}

\item{-} The NCAdS background \background, \RRbackground\ has a generalization to $D=4-2\epsilon$ dimensions \AlishahihaCI. The dimensionally regularized background is obtained from \background\ by simply replacing $u^2$ with $u^{2+\epsilon}$.

\subsec{T-dual of planar non-compact NCYM} 

By performing disk level T-duality along the bosonic transverse directions \AldayHR\ for the background \background\ we find the background 
\eqn\Tdualbackground{\eqalign{ds^2=&{d u^2\over u^2}+{1\over u^2}\(-dy_{-1}^2-dy_0^2+dy_1^2+dy_2^2\)\cr=&{dr^2\over r^2}+ r^2\(-dy_{-1}^2-dy_0^2+dy_1^2+dy_2^2\)~,\cr\cr \tilde B_{-1\,0}=&\tilde B_{1\,2}=\theta={1\over B_\infty}~,\qquad e^{\phi}={g\over u^4}~,}}
where $r={1\over u}$. That is an $AdS_5$ background with a constant NS $B$-field. We expect that an addition fermionic T-duality will remove the running dilaton \BerkovitsIC. The constant addition to the NS $B$-field will not enter the type IIB supergravity equations (as $C_0$ is constant) and couples only to the boundary of open strings. Near the AdS boundary, the metric diverges and therefore the constant $B$-field is negligible. Near the Poincare horizon, the metric shrinks to zero and is therefore negligible with respect to the constant $B$-field.

\subsec{The NCAdS dual of Wilson loops and scattering amplitudes}

Consider an open string in $AdS_5$ with Dirichlet boundary conditions in the radial direction. There are two consistent boundary conditions one can impose in the transverse directions. One is Dirichlet boundary condition and the other are Neumann boundary condition. The field theory Wilson loop along the polygon \polygon\ is dual to imposing Dirichlet boundary condition in the transverse direction along the loop \refs{\MaldacenaIM, \AldayHR}. The radial position of the open string then plays the role of a UV cutoff. More correctly, the field theory size of the loop measured with respect to the UV cutoff scale is identified with the proper size of the loop on the open string boundary measured in $R_{AdS}$ units. The gluon scattering amplitude is dual to imposing Neumann boundary conditions and in addition to inserting gluon vertex operators ordered along the open string boundary \AldayHR. Similarly, the radial position of the open string plays the role of an IR cutoff.

Consider now an open string in the NCAdS background \background\ with Dirichlet boundary conditions in the radial direction ($u$). Again, there are two consistent boundary conditions one can impose in the transverse directions. One is Dirichlet boundary condition and the other is the mixed Neumann-Dirichlet boundary condition
\eqn\boundarycond{\[G_{\mu\nu}(u)\d_nx^\nu+2\pi iB_{\mu\nu}(u)\d_tx^\nu\]|_{\d\Sigma} =0~.}
Therefore, in the limit where $G$ is negligible with respect to $B$, the two consistent boundary conditions coincide.

As we will see below, the non-commutative scattering amplitude \convervedamp\ is dual to imposing the boundary conditions \boundarycond\ in the transverse directions and in addition inserting the ordered gluon vertex operators along the open string boundary. The radial position of that open string is dual to the field theory IR cutoff. Surprisingly, the non-commutative Wilson loop \conservativeWilson\ is dual to doing the exact same thing in the shifted background \shifted. That is, also imposing the boundary conditions \boundarycond\ in the transverse directions and adding vertex insertions ordered along the open string boundary. But with a $B$-field that is shifted with respect to the one in \background\ and an open string that is extended in the opposite radial direction. Now, the radial position of that open string is dual to the field theory UV cutoff.

\subsec{The NCAdS dual of scattering amplitudes}

The backgrounds \background\ and \Tdualbackground\ are related by disk level T-duality. Suppose we start with an open string in \background\ with the mixed Neumann-Dirichlet boundary conditions \boundarycond\ in the transverse directions and gluon vertex insertions ordered along its boundary. Under the disk-level T-duality, it maps to a T-dual open string in the background  \Tdualbackground. The T-dual open string has Dirichlet boundary conditions in the radial direction (as we T-dualize the transverse directions only) as well as Dirichlet boundary conditions in the transverse directions along the polygon loop \polygon. As the background \Tdualbackground\ is a pure AdS background with a constant NS $B$-field, the resulting open string observable equal to the ordinary planar polygon Wilson loop. Through the Alday-Maldacena duality, that is also the same as the ordinary planar gluon scattering amplitude. The constant NS $B$-field only gives an overall dressing phase equal to the $B$-field flux through the loop
$$\Phi=\half\int  d\sigma \tilde B_{\mu\nu}x^\mu\dot x^\nu=\half\int  d\sigma\theta_{\mu\nu}x^\mu\dot x^\nu~.$$
For the polygon loop \polygon, that phase equals
\eqn\phase{\Phi=\half\sum_{i<j}k_i^\mu\theta_{\mu\nu} k_j^\nu~.}
Therefore, we have reproduced in the strong coupling limit the perturbative result \planarrelation\ \FilkDM. In particular, it confirms that the radial regulator is compatible with non-commutativity and that the identification of $u$ with the gauge theory energy scale is not corrected by the presence of the non-commutativity scale. 

Note that the classical solution embedded in the NCAdS background  \background\ always extends from $u_{IR}$ in the direction of the boundary.\foot{In dimensional regularization for example, the region $u<u_{IR}$ does not exist.} For large $\theta u_{IR}^2\gg1$ the $B$-field is constant and the metric becomes AdS again, but now it is shrinking in the $u$-direction:
\eqn\asymptoticAdS{\eqalign{&ds^2\to{du^2\over u^2}+{dx_{(2,2)}^2\over \theta^2 u^2}={du^2\over u^2}+{d\tilde x_{(2,2)}^2\over  u^2}\cr &B_{\tilde x\tilde x}\to\theta~,\qquad e^\phi\to{g\over\theta^2u^4}~,}}
where $\tilde\bx=\bx/\theta$. In that region, the fact that the classical solution for the scattering amplitude extends in the direction where the metric shrinks is somewhat counter-intuitive. The reason for that is the presence of a large NS $B$-field (compared to the metric). It causes the gluon open string states to carry \boundarycond\ ``winding" and therefore, as $\theta u_{IR}^2$ becomes large, the problem looks more and more like the polygon Wilson loop calculation. In other words, {\it the region where $\theta u^2\gg1$ describes the physics in terms of the T-dual variables!} In this region, T-duality is induced by the large $B$-field.\foot{Note that in dimensional regularization, the sign of $\epsilon=(4-D)/2$ in the $\theta u^2\gg1$ AdS throats is the same as in the T-dual AdS.} The NCAdS background can be thought of as describing the same physics as a pure AdS background, but in terms of variables that are rotated between position space and momentum (loop) space in a scale dependent way.

If $\theta u_{IR}<1$ but $\bk_i\cdot\bk_{i+1}/u_{IR}^2\gg1$, the classical scattering amplitude solution will penetrate into the $\theta u^2\gg1$ region. In that region, it will look like the polygon Wilson loop solution. That is an outcome of the region near the NCAdS boundary being T-dual to itself, without inversion of the radial coordinate. To see that we start from the background \Tdualbackground, T-dualize back to NCAdS and focus on the $\theta u^2\gg1$ region:
$$\d_\alpha \tilde x^\mu=\[ig_{\mu\nu}(u)\epsilon_{\alpha\beta}\d_\beta y^\nu+B_{\mu\nu}\d_\alpha y^\nu\]/\theta\rightarrow B_{\mu\nu}\d_\alpha y^\nu/\theta=M_{\mu\nu}\d_\alpha y^\nu~.$$

If we now hold $u_{IR}$ fixed and take $\theta$ to be large such that $\theta u_{IR}^2\gg1$, the scattering amplitude problem becomes equivalent to a ``scattering problem" in a pure AdS background with constant $B$-field equal to $\theta$ \asymptoticAdS\ and gluon momenta equal to $\theta\bk_i$. In addition, the open string has to extend in the direction where the AdS metric shrinks ($u>u_{IR}$). That is the opposite direction to the one relevant for scattering amplitudes (in \asymptoticAdS) with zero $B$-field. We do not know how to solve that problem in AdS.\foot{For $u_{IR}=0$, the solution for the four gluon amplitude can be obtained from the one in AdS \AldayHR\ by first turning on a constant $B$-field and then T-dualizing the result.} To gain a better understanding, it is instructive to repeat the Gross-Manes flat space scattering amplitude \GrossGE\ in the presence of a constant background NS $B$-field and rescaled momenta. We define
$$B_{\mu\nu}=bM_{\mu\nu}~,\qquad\({1\over g+ B}\)_{\mu\nu}=G_{\mu\nu}+\theta_{\mu\nu}~,\qquad k^{(i)}_\mu=B_\mu^\nu p^{(i)}_\nu~,$$
where $M$ is given in \Mmatrix, $g$ is the flat metric in $(2,2)$ signature, $G$ is symmetric, $\theta$ is anti-symmetric and we concentrate on the four gluon amplitude $i=1,2,3,4$. The classical solution is given by
$$x^\mu(z,\bar z)=i\sum_{i=1}^4\[ G^{\mu\nu}k^{(i)}_\nu\log|z-\sigma_i|+\theta^{\mu\nu}k^{(i)}_\nu\log\({z-\sigma_i\over\bar z-\sigma_i}\)\]~,$$
where $|\bk^{(i)}|^2=0$ and $\{\sigma_i\}$ are the vertex operators insertion points. These are related to the external gluon momenta by
$${(\sigma_1-\sigma_3)(\sigma_2-\sigma_4)\over (\sigma_1-\sigma_3)(\sigma_2-\sigma_4)} =-{(\bk^{(1)}+\bk^{(2)})^2\over (\bk^{(1)}+\bk^{(3)})^2}= -{s\over t}~.$$
The on-shell worldsheet action is given by
\eqn\flatact{S_{flat}=-s\log s-t\log t+(s+t)\log(s+t)+{i\over 2}\sum_{i<j}k_\mu^{(i)}\theta^{\mu\nu}k_\nu^{(j)}~,}
where
$$s=k_\mu^{(1)}G^{\mu\nu}k_\nu^{(2)}~,\qquad t=k_\mu^{(1)}G^{\mu\nu}k_\nu^{(3)}~.$$

Next, we take the $b\to\infty$ limit while holding the $\bp^{(i)}$'s fixed. The resulting classical solution and amplitude are given by
\eqn\limitedsol{x^\mu(z,\bar z)=i\sum_{i=1}^4\[ {M^{\mu\nu}\over b}p^{(i)}_\nu\log|z-\sigma_i|+p^{(i)}_\nu\log\({z-\sigma_i\over\bar z-\sigma_i}\)\]~,}
$$S_{flat}=-\tilde s\log\tilde s-\tilde t\log\tilde t+(\tilde s+\tilde t)\log(\tilde s+\tilde t)+{i\over 2}\sum_{i<j}p_\mu^{(i)} B^{\mu\nu}p_\nu^{(j)}~,$$
where
$$\tilde s=p^{(1)}_\mu p^{(2)}_\nu g^{\mu\nu}~,\qquad \tilde t=p^{(1)}_\mu p^{(3)}_\nu g^{\mu\nu}~.$$
Apart from the phase, that is exactly the flat space result for the scattering amplitude with momenta $\bp^{(i)}$'s as well as the T-dual Dirichlet problem with polygon of edges $\bp^{(i)}$. 

The first term in \limitedsol\ is proportional to $1/b$. Therefore, for any value of $z$ different from one of the insertion points ($\sigma_i$), it is negligible with respect to the second term in \limitedsol. If we subtract the first term from \limitedsol\ we find the flat space solution for the Dirichlet problem with polygon of edges $\bp^{(i)}$. However, no matter how big $b$ is, close enough to the insertion points the first term in \limitedsol\ dominates. It dresses each edge of the real polygon with an infinite imaginary arm in the $M\bp^{(i)}$ null direction. If we cut out a small region of the worldsheet around each gluon insertion, then we can neglect the first term in \limitedsol\ and the solution reduces to the polygon one. In addition, the on-shell action (including the $B$-field coupling and the vertex insertions) remains the same.\foot{In the $b\to\infty$ limit, the first term in \limitedsol\ has two contributions to the on-shell action. One from the $B$-field coupling contracted with the second term in \limitedsol\ and the other from the vertex insertions. These two contributions cancel each other.}  Therefore, in the $b\to\infty$ limit, the infinite imaginary null arms do not contribute to the classical action. Cutting off a small region around the vertex insertion is a sort of UV regulator on the world sheet. 

Again, in AdS with constant NS $B$-field we don't know how to solve the classical problem. However, we expect the rough picture to be the same. That  is, the classical solution will look like the real polygon with imaginary null arms attached to each edge. In the $\theta u_{IR}^2\gg1$ limit, the imaginary arms will not contribute to the on-shell action.\foot{Now, as we try to decrease $u_{IR}$ within the $\theta u_{IR}\gg1$ region, we get a large contribution to the on-shell action, not because there is an infinite arm that goes all the way to infinity, but because the proper size of the edges becomes large.} In that limit, if we cut the solution at $u=u_{IR}+\epsilon$, then for small $\epsilon$ we will find the polygon Wilson loop solution. In a sense, that is an outcome of the region near the NCAdS boundary been T-dual to itself, without inversion of the radial coordinate.

We interpret these imaginary arms dressing the polygon edges as an analog of a field theory picture in which we dress each edge of a polygon Wilson loop by an amputated on-shell gluon. The infinite arms represent the amputation. This naive picture suffers from the problem that the on-shell gluon is in momentum space whereas the polygon Wilson loop is in position space. The picture will become clearer in the section 4, where we will study in field theory the scattering amplitude in the large magnetic background. 

In the next section we will see that the above picture also reveals the NCAdS dual of Wilson loops.

\subsec{The NCAdS dual of Wilson loops}

What is the NCAdS dual of the Wilson loop (2.4)? As for Wilson loops in OYM, we expect it to be described by an open string in NCAdS with boundary conditions specified at the NCAdS boundary (or at $u_{UV}$ in radial regularization). In addition, at the $\theta u^2\ll1$ region, the classical string solution should reduce to the ordinary one in pure AdS. In \MaldacenaMH\ it was shown that one cannot impose Dirichlet boundary conditions in the transverse directions at the NCAdS boundary. Finally, in section 2 we learned that the planar (disk level) result should be completely independent of the non-commutativity scale. 

In the previous subsection we saw that the region $\theta u^2\gg1$ (where the NCAdS boundary is) describe the physics in terms of the T-dual variables. Given these observations, the answer is clear: we should consider the same open string as described above, but in the coordinates \shifted. That is, we should first shift the NS $B$-field by a constant ($B_\infty$), such that it will vanish at the boundary ($u\to\infty$).\foot{Note that after the constant shift in the $B$-field, the region $\theta u^2\gg1$ is the same as the T-dual of pure AdS with $g_s$ rescaled by $\theta^2$. For the running dilaton to agree, it is crucial that the T-duality at hand is only the bosonic one \BerkovitsIC.} Then, we should consider an open string scattering amplitude with gluon momentum equal to $\bk_i/\theta$, imposing Neumann boundary conditions in the transverse directions and Dirichlet boundary condition in the radial direction at the NCAdS boundary (or at large $u_{UV}$ in radial regularization, with mixed Neumann-Dirichlet boundary conditions \boundarycond\ in the transverse directions). The corresponding classical solution now extends in the direction where $u$ decreases. Near the NCAdS boundary, it looks like the ordinary scattering amplitude solution in AdS.\foot{As a side remark, we note that when perpendicularly approaching the open string boundary of the Alday-Maldacena solution for the four gluon scattering amplitude \AldayHR, we find a linear relation $\bx\propto u$, in agreement with the observation in \MaldacenaMH.} If the classical solution penetrates into the ordinary region $\theta u^2\ll1$, it will look as the ordinary Wilson loop there. As before, by applying T-duality we find an open string describing the Wilson loop in the pure AdS background with constant NS $B$-field \Tdualbackground. However, now $r=\theta u$ (in radial regularization $r_{UV}=\theta u_{UV}$) and the size of the loop in the $\by$-coordinate is rescaled by $1/\theta$. These two rescalings cancel each other. The constant $B$-field dresses the result by a phase equal to the $B=\theta$ flux through the polygon loop \polygon\ rescaled by $1/\theta$:
$$e^{i\Phi}=\exp\[{i\over 2}\sum_{i<j}k_i^\mu\({1\over\theta}\)_{\mu\nu} k_j^\nu\]~.$$
Therefore, the result should be divided by that phase factor.

Again, we have reproduced at strong coupling the perturbative result \Wilsoninvariant\ \FilkDM. The prescription has a natural generalization to non-polygon smooth loops. As in \AldayHE, these can be approached by a zig-zag contour and are therefore described by a gluon scattering amplitude at the NCAdS boundary with infinitely many soft gluons and shifted $B$-field \shifted.

In the next section we go back to the field theory and study the planar scattering amplitude in the presence of a large background magnetic field.

\newsec{Scattering amplitude in a large magnetic background - the field theory side}

We saw in section 3.4 that if one studies the scattering amplitude using radial regularization and take the non-commutativity scale to be much larger than the IR cutoff ($\theta u_{IR}^2\gg1$) then the dual open string in NCAdS (that extends in the $u\ge u_{IR}$ direction), lives in the $\theta u^2\gg1$ region. In that region, the NCAdS background becomes AdS with constant NS $B$-field \asymptoticAdS\ and the scattering amplitude looks like a polygon Wilson loop computation. That region can be thought of as the dual of planar $\NN=4$ SYM in a large magnetic background. The scattering amplitude in that background has to be computed with a UV regulator (instead of an IR regulator). In this section, we will do that computation directly in the gauge theory. We will find that in the limit of large magnetic background, the scattering amplitude computation reduce to the computation of a complex polygon Wilson loop expectation value (now UV regulated) dressed by the phase \phase\ and by a polarization dependent factor. 

To prepare the ground for the calculation, we start from a preliminary discussion of Wilson loops in a large magnetic background. That is the standard Wilson loop computation, for which the addition of the large magnetic background is trivial.

\subsec{Wilson loops in large magnetic background}

In section 3.4 we saw that the planar scattering amplitude in the NCAdS background \background\ is T-dual to a polygon Wilson loop in a constant NS $B$-field background \Tdualbackground. What is the dual of a constant NS $B$-field in AdS?
To answer this question lets first consider in more details the AdS dual of the polygon Wilson loop with no NS $B$-field. In the radial regularization we work with here, the open string in AdS has Dirichlet boundary conditions in the radial AdS direction. The radial position of the open string boundary plays the role of a UV regulator. In the field theory dual, the Wilson loop is the phase associated to a probe quark that propagates along the loop. The radial position of the open string boundary is dual to the mass of the corresponding probe quark. Removing the UV regulator corresponds to the limit where the quark mass is sent to infinity. The quark represents a massive probe in the fundamental representation. A way to realize such massive fundamental probe in $\NN=4$ SYM in the planar limit is to first go on the Coulomb branch, breaking an $SU(N+n)$ gauge group into $[SU(N)\times SU(n)]\times U(1)_{diag}$ gauge group. Then taking the 't Hooft large $N$ limit, while keeping $n$ and $\lambda=g_{YM}^2N$ fixed (for simplicity one may take $n=1$). In that planar limit, the $SU(n)$ gauge group becomes free and may be thought of as a global symmetry. Now, any field in the bi-fundamental of $SU(N)\times SU(n)$ is in particular a massive field in the fundamental of $SU(N)$. The propagation of such field between two points will result in the corresponding regularized Wilson line holonomy.

Turning on a constant NS $B$-field in AdS dress the Wilson loop by the $B$-flux through the loop. To reproduce that coupling in the field theory, the fundamental probe field should be coupled to a constant magnetic background. For the field in the bi-fundamental of $SU(N)\times SU(n)$ to be charged under a $U(1)$ magnetic background, the latter must be in the $U(1)_{diag}$ of $[SU(N)\times SU(n)]\times U(1)_{diag}$. Moreover, the coupling of the bi-fundamental field to the magnetic flux is proportional to $g_{YM}$. Therefore, in order for that coupling to survive the 't Hooft limit, we must simultaneously scale the magnetic background with ${1\over g_{YM}}=\sqrt{N\over\lambda}$. 

We conclude that in the field theory dual one has
\item{1)} To go on the Colombo branch breaking $SU(N+n)\to [SU(N)\times SU(n)]\times U(1)_{diag}$.
\item{2)} To consider the gauge theory in a state where there is a constant magnetic flux in the $U(1)_{diag}$. The field theory magnetic flux $b_{\mu\nu}$ is related to the NS $B$-field by $g_{YM}b=B$. In the field theory language, at the planar limit, the magnetic flux is scaled to infinity.
\smallskip
In the non-compact background we are working in, a state supporting such a constant magnetic flux is not normalizable and has infinite energy. Therefore, we should shift the vacuum energy correspondingly and think about that state as the new vacuum of the shifted theory.

\subsec{Scattering amplitudes in a large magnetic background}

What is the the dual of an AdS open string scattering amplitude in a large constant NS $B$-field (and UV regulator)? Given our experience from the previous subsections, the answer is clear. One has to break the gauge group as $SU(N+n)\to[SU(N)\times SU(n)]\times U(1)_{diag}$ and turn on large magnetic field in the $U(1)_{diag}$ part. The external gluons live in the (probe) $SU(n)$ part and carry momentum that is proportional to the large magnetic field. The result also has to be UV regulated.

To set the ground for the computation, we list the steps below:
\item{1)} Consider an $SU(N+n)$ gauge group. 
\item{2)} Go on the Coulomb branch, breaking it to $SU(N+n)\to[SU(N)\times SU(n)]\times U(1)_{diag}$ and giving the $w$-bosons a mass $m$. 
\item{3)} Turn on constant background magnetic field $b_{\mu\nu}=bM_{\mu\nu}$ in the $U(1)_{diag}$ part.
\item{4)} Take the $N\to\infty$ limit while holding $n$, $\lambda=g_{YM}^2N$ and $B=g_{YM}b$ fixed.
\item{5)} In the limit (4) consider an $SU(n)$ partial gluon scattering amplitude of momenta $\bk_i$, $i=1,\dots,m$, $m<n$.
\item{6)} Add a UV regulator.
\item{7)} Take the limit $B\to\infty$ while holding $p_i^\mu={\({1\over B}\)^\mu}_\nu k_i^\nu$, the UV regulator and the physical mass of the $w$-bosons ($m_{phys}$) fixed. That small physical mass will not enter the final result and can be set to zero.
\item{8)} Multiply the result by the overall phase $\exp\{-{i\over 2}\sum_{i<j}p_i^\mu B_{\mu\nu} p_j^\nu\}$.

\smallskip\smallskip


Only the fields in the bi-fundamental of $SU(N)\times SU(n)$ are charged under the $U(1)_{diag}$ and therefore affected by the magnetic field. As a result, the $SU(n)$ tree level diagrams are not affected by the magnetic field. The fields in the adjoint of $SU(N)$ run at the interior of loop diagrams and are not affected by the magnetic field as well. The fields in the bi-fundamental of $SU(N)\times SU(n)$ run along the boundary of planar loop diagrams (for more details see \McGreevyZY). They are charged under the $U(1)_{diag}$ part and therefore their propagator depends on $B$. There are two different ways of studying the dynamics of the bi-fundamental fields at large magnetic background, both lead to the same result. One is by writing the bi-fundamentals propagator explicitly. There is a pole in that propagator for each Landau level. In the $B\to\infty$ limit, the pole from the first Landau level dominate. Restricting to it leads to the desired result. We refer the reader to the relevant literature on the subject \refs{\LLL, \SalimKY}. Here instead, we choose a more intuitive approach (also considered in \SalimKY) expressing the bi-fundamental propagation in the worldline formalism. The resulting worldline particle is charged under the background magnetic field.

\smallskip\smallskip\smallskip\smallskip

{\bf Planar gluon scattering amplitude at strong magnetic background in the worldline formalism}

\smallskip\smallskip

At the planar limit, the fields in the bi-fundamental run only on the boundary of planar diagrams and contribute only at one loop to 1PI diagrams \McGreevyZY. These one loop contributions can be represented in the worldline formalism \StrasslerZR. Here, for simplicity we will concentrate on the scalar contribution only, given by \ScattWilson. The worldline coupling to the $SU(N)\times SU(n)$ gauge fields $(\bA,\ba)$ is \McGreevyZY
$$S[\bA,\ba]=ig_{YM}\int ds\,\dot\bx\cdot\[\bA-\ba\]~.$$
The $U(1)_{diag}$ is generated by
\eqn\diaggen{T_{diag}={1\over\sqrt{N^2n+n^2N}}\(\matrix{N\One_{n\times n}&\cr&-n\One_{N\times N}}\)\quad\to\quad {1\over\sqrt n}\(\matrix{\One_{n\times n}&\cr&0}\)~.}
Therefore, turning on the background magnetic field adds to the worldline action the term\foot{Where we absorbed the $\sqrt n\sim 1$ of \diaggen\ in $B$.}
$$S_B={i\over 2}\int ds\,B_{\mu\nu}\dot x^\mu x^\nu~.$$
That is the coupling of a pointlike particle to a background magnetic field. The worldline action becomes 
\eqn\worldlineaction{S=\int_0^T ds\[\half \dot\bx^2+m^2+ig_{YM}\bA\cdot\dot\bx+{i\over 2}B_{\mu\nu}\dot x^\mu x^\nu+i\sum_i\bk_i\cdot\bx\,\delta(s-s_i)\]~,}
where the last term accounts for the gluon momentum insertions. 

Next, we first turn off the worldline coupling to the gauge field (given by the corresponding Wilson loop expectation value). We will argue that now, in the strong magnetic background limit, the worldline path integral is dominated by a saddle point. Fluctuations around that saddle point are exponentially suppressed in the strong magnetic limit. Turning back the coupling to the (UV regulated) Wilson loop will not change that picture.

The worldline equation of motion is
$$\ddot x^\mu(s)-i{B^\mu}_\nu\dot x^\nu=i\sum_i k_i^\mu\delta(s-s_i)=i\sum_i{B^\mu}_\nu p_i^\nu\delta(s-s_i)~.$$
The solution is given by\foot{The general solution to a second order differential equation has two integration constants. Here, one of these constants is determined by demanding that the contour form a closed loop.}
\eqn\claswlssol{\bx(s)=-{iT\over 4\ZZ}\sum_i\[{e^{-i\dot G_B(s,s_i)\ZZ}\over\sin\ZZ}+i\,{\rm sign}(s-s_i)\]\bk_i+\bx_0~,}
where $\bx_0$ is an integration constant, $\ZZ_{\mu\nu}={T\over 2}B_{\mu\nu}$ and
$$\dot G_B(s,s_i)={\rm sign}(s-s_i)-2{s-s_i\over T}$$
is the derivative of the worldline Green function. Note that the function \claswlssol\ is smooth and periodic. Since the matrix $i\ZZ$ is hermitian and $|\dot G(s,s_i)|<1$, in the $TB\to\infty$ limit the solution \claswlssol\ reduce to
\eqn\limitedpolygon{\bx_{polygon}(s)~\to~{1\over B}\sum_i\bk_i\theta(s-s_i)+\bx_0=\sum_i\bp_i\,\theta(s-s_i)+\bx_0~.}
That is our polygon loop. It is periodic, but no longer smooth. The same solution is obtained if one naively neglect the kinetic term with respect to the magnetic field coupling in the equation of motion. The approximation \limitedpolygon\ is valid at any point on the worldline except at the insertions points $\{s_i\}$. At an insertion point $\dot G(s_i,s_i)=0$ while $\dot G^2(s_i,s_i)=1$. As a result
$${e^{-i\dot G_B(s_i,s_i)\ZZ}\over\sin\ZZ}=\cot\ZZ\to{1\over M}~.$$
We now find
\eqn\insertionpoint{\bx(s_i)\to\[{1\over B}\sum_{j<i}\bk_j+\bx_0\]+{i\over 2|B|}\bk_i=\[\sum_{j<i}\bp_j+\bx_0\]+{i\over 2} M\bp_i~.}
Therefore, the $(\bp_i)$ edge of the real polygon \limitedpolygon\ is replaced by the two adjacent edges $\half(1+iM)\bp_i\to\half(1-iM)\bp_i$. The classical solution is now a complex polygon made of $2m$ edges. We will call these imaginary shifts of the real $\bp_i$ edge - ``imaginary arm". These are the worldline analog of the string theory imaginary arms in \limitedsol. 

When we now plug the classical solution \limitedpolygon\ back into the action \worldlineaction, we must include the imaginary point-like arms \insertionpoint\ or alternatively, use the full solution \claswlssol\ and only at the end take the $TB\to\infty$ limit. Doing so, we obtain following limited on shell action\foot{Note that the imaginary arms are responsible for the important last term.}
\eqn\onshellworldline{\eqalign{S_{\rm on-shell}\to&{i\over 2}\sum_{i<j}k_i^\mu\({1\over B}\)_{\mu\nu}k_j^\nu-{1\over 2|B|}\sum_i\bk_i^2\cr =&{i\over 2}\sum_{i<j}p_i^\mu B_{\mu\nu}p_j^\nu-\half |B|\sum_i\bp_i^2 ~.}}

In the $TB\to\infty$ limit, fluctuations around the saddle point are exponentially suppressed. The determinant of these fluctuations leads to an exponentially suppressing factor
$${e^{-m^2T}\over T}{\det'}_T^{-\half}(-D^2+m^2)\quad\to\quad{|B|\over T^2}e^{-(m^2+|B|)T}~,$$
where $\det'_T$ is the worldline determinant excluding the zero modes, $Dx^\mu=\d_sx^\mu+{i\over 2}{B^\mu}_\nu x^\nu$ is the worldline covariant derivative in a magnetic background and we included the worldline mass dependence factor ${e^{-m^2T}\over T}$ in \ScattWilson. The worldline representation of one loop determinants rely on the Schwinger representation
\eqn\Schwinger{\log(a)=\int_0^\infty{dT\over T} e^{-aT}~.}
The integral in \Schwinger\ is actually divergent and what one means by \Schwinger\ is that in the limit $\epsilon_{UV}\ll1/a$
$$\int_{\epsilon_{UV}}^\infty{dT\over T} e^{-aT}\quad\to\quad\log(a/\epsilon_{UV})~,$$
where $\epsilon_{UV}$ is a UV cutoff scale. The leading contribution to \Schwinger\ is therefore from $T\in[\epsilon_{UV},{1\over a}]$. In our case 
$$a=m^2+|B|~.$$
The $|B|$ addition to the worldline mass is the shift of the worldline vacuum energy due to the energy in the first Landau level.

For \Schwinger\ and our approximation to be valid, we must tune the bare mass with $B$, such that the physical mass and the UV cutoff are kept fixed in the $B\to\infty$ limit. It is important that the on-shell worldline action \onshellworldline\ is independent of the bare mass. At the end, one may set the physical mass to zero. 

The first term in \onshellworldline\ is the overall dressing phase we divide by (that is true both, for 1PI and for non-1PI contributions \FilkDM). If the momentum insertions along the loop are the on-shell external ones and the insertion points do not coincide ($s_i\ne s_{i+1}$), then $\bp_i^2=0$ and the second term in \onshellworldline\ is zero. If the momentum insertions along the loop are not the external ones, then the full planar diagram at hand is not 1PI. In that case, the momentum insertions along the loop $\bk_i$ are sums of adjacent external gluon momenta. As the external momenta scale as $B$, so are the momentum insertions along the loop $\bk_i=B\bp_i$. Since we consider the case of only space-like momentum transfer in all channels, $\bp_i^2>0$ (and is held fixed as $B\to\infty$). The resulting contribution is therefore exponentially suppressed and vanishes as $B\to\infty$.\foot{An exception are bubble 1PI diagrams dressing external on-shell gluons propagator. As the $SU(n)$ gauge group is not broken, these cannot give a mass to the external gluons, whose on-shell external propagator is amputated.} If the insertion points of two (adjacent) external gluons coincide ($s_i=s_{i+1}$) then the total external momentum insertion at that point is space-like. These coincident insertion points correspond to the case where two external gluons couples to the w-boson in a four vertex. As for the non-1PI contributions, the result is exponentially suppressed. The large magnetic field then gives a projection to the 1PI planar diagrams in which the external gluons cupels to the w-bosons in a three vertex only.

In addition, the saddle point is dressed by the gluon polarization dependent weights in \ScattWilson\ ($\epsilon_i\cdot\dot\bx(s_i)$) and we must check that these do not vanish. The first derivative of \claswlssol\ is 
$$\dot\bx(s)=\half\sum_i{e^{-i\dot G_B(s,s_i)\ZZ}\over\sin\ZZ}\bk_i\quad\to\quad\sum_i{\delta_{s,s_i}\over 2M}\bk_i=\half|B|\sum_i\delta_{s,s_i}\bp_i~.$$
Plugging it into the polarization dependent weights we find
\eqn\polarizations{\epsilon_i\cdot\dot\bx(s_i)=\half|B|\epsilon_i\cdot\sum_j\bp_j\delta_{s_i,s_j}~.}
When integrating over the ordered insertion points, the Kronecker delta functions $\delta_{s_i,s_j}$ contribute only if $i=j$. The resulting polarization dependent dressing factors are
$$\epsilon_i\cdot\dot\bx(s_i)=-\half\epsilon_i\cdot M\bk_i=\half|B|\epsilon_i\cdot\bp_i~.$$

Let 
$$\bk=(0,1,0,1)$$
be a null momentum in $(2,2)$ signature. The corresponding two physical (real) polarization vectors are
$$\epsilon^+=(1,0,1,0)~,\qquad \epsilon^-=(1,0,-1,0)~.$$
Since 
$$M\bk=(1,0,1,0)$$
we have\foot{Note that the longitudinal polarization decouple.}
$$\epsilon^+\cdot M\bk=2~,\qquad \epsilon^-\cdot M\bk=0~.$$
We find that the polarization dependent dressing factor is non-zero only for amplitudes where all polarizations are $+$'s. That is a point-like version of the Zeeman effect, where different polarizations have different energies .\foot{We thank J. Maldacena for point that to us.}

To summarize, the strong magnetic field together with the rescaling of the gluon momenta have four effects:
\item{1)} It turns the gluon momentum insertions on the worldline into ``winding" insertions.
\item{2)} Between the gluon insertions, the worldline particle is projected to the first Landau level where it essentially becomes localized. That is the dual of the string theory picture, where Neumann boundary conditions are changed into Dirichlet \boundarycond.
\item{3)} It projects out all non-1PI diagram, leaving the 1PI contributions to the gluon amplitude in which the external gluons cupels to the w-bosons in a three vertex only (and the tree level ones).
\item{4)} It projects out all amplitudes where not all polarization are $+$'s. 

What happens if we now turn back the coupling to the gauge field?
Turning on the coupling to the gauge field dresses the worldline action by the Wilson loop expectation value
\eqn\Wilsonexpectation{\<W[\bx(\cdot)]\>~.}

The Wilson loop is reparametrization-invariant and only depends on the loop image. The image of our classical loop \claswlssol\ has imaginary ``arms". So what do we mean by a Wilson loop along a complex contour? In perturbation theory the answer is clear: Compute the gauge field correlation functions in momentum space and analytically continue the Fourier transform coefficients.

Since the classical solution is independent of the magnetic field ($B$), so is the Wilson loop expectation value. As long as there are no divergences, in the $B\to\infty$ limit it cannot push the worldline particle away from the saddle point above (that is, away from the first Landau level). Moreover, the Wilson loop is UV regulated and the regulator is held fixed in the $B\to\infty$ limit. Therefore there cannot be any dangerous divergences.

The Wilson loop expectation value has two types of UV divergences. One is from mass renormalization diagrams and the other is from cusps. We now add a UV regulator and hold it fixed in the $B\to\infty$ limit. The first UV divergence, leads to renormalization of the worldline particle mass \PolyakovCA. As before, it is accounted for by tuning the bare worldline mass such that the physical mass in the effective worldline action is finite (and can be set to zero). The second divergence, coming from cusps, is logarithmic. In the limit considered here, where the UV cutoff is held fixed while $B\to\infty$, that logarithmic contribution cannot affect the saddle point.

The same picture is seen in perturbation theory. When expanding \Wilsonexpectation\ in perturbation theory, we are inserting some finite number ($m$) of ordered integrated gauge fields along the loop. These internal gauge fields are carrying the integrated momenta $\widetilde\bk_i^{(j=1,\dots,m_i)}$, which are new sources for the worldline particle:
\eqn\Fourier{i\Psi=i\sum_{i=1}^n\sum_{j=1}^{m_i}\widetilde\bk_i^{(j)}\cdot\bx(s_i^{(j)})}
dressed by the trace of the ordered gauge fields correlation function. Here, $\widetilde\bk_i^{(j)}$ is the momentum of the $j$'th internal gluon inserted between the $i$'th and the $(i+1)$'th external gluons. Repeating the above analysis, result in a polygon with new edges. These edges are of length
$$\Delta x^\mu(s_i^{(j)})=\({1\over B}\)^{\mu\nu}\(\widetilde\bk_i^{(j)}\)_\nu~.$$
In the large magnetic field limit they are negligible with respect to the external edges. The internal momenta also enters the on-shell worldline action \onshellworldline. That result in three new terms. Two of these terms scale as $1/B$ and therefore, for any fixed internal momenta (bounded by the UV cutoff), vanishes as $B\to\infty$. The only term that survive is
\eqn\cuspphase{S_{internal}=i\sum_{i,j}\[\widetilde\bk^{(j)}_i\cdot\sum_{l<i}\bp_l\]~.}
That is a Fourier transform phase, dressing the gauge fields correletion function and putting the internal gluons at the polygon loop cusps \limitedpolygon. If however, the insertion point of an internal gluon coincide with the insertion point of an external gluon, then \cuspphase\ receive an addition contribution now placing the internal gluon at the corresponding imaginary point \insertionpoint. 

We conclude that in the infinite magnetic field limit, the scattering amplitude reduces to a polygon Wilson loop computation with the imaginary arms \insertionpoint, dressed by a polarization dependent factor. In addition, one should add the $SU(n)$ tree level contributions that are not effected by the $B$-field.

In section 3, when studying the string theory picture, we argued that the imaginary arms do not contribute in the $\theta\to\infty$ limit and can be removed, leaving the real polygon loop \limitedpolygon\ instead. One way of seen that was by cutting off the open string ending on the probe brane in the $\theta u_{IR}^2\gg1$ region of NCAdS \background\ at some small redial distance $u=u_{IR}+\epsilon$ from the brane. Another way was by studying the flat space scattering amplitude in a large magnetic background. There, we saw that cutting off a small region around each vertex insertion on the worldsheet removes the imaginary arms without changing the on-shell action. If true, the presence of the imaginary arms should not affect the regulated Wilson loop expectation value. However, we where not able to show that directly in the field theory calculation done here. We leave the study of the complex polygon Wilson loop for future works and make one last speculation: 

In supersymmetric gauge theories the first non-zero amplitudes are the MHV ones. It is possible to turn on a $B$-field without breaking some supersymmetry. In this case, also in the presence of the $B$-field, we expect the first non-zero amplitudes to be the MHV ones. For $\NN=4$ SYM, we expect the polarization dependent factor to equal the tree-level amplitude. If true, one can also use our formalism to understand the polarization dependent dressing factor for non-MHV amplitudes \Inprogress.

\newsec{Closed strings in NCAdS and the dual of single trace correlation function}

In this paper we saw that the NCAdS dual of Wilson loop has to be specified at the NCAdS boundary in terms of the T-dual variables and a shifted $B$-field. Here, we would like to suggest that the same is true for closed string observables.\foot{An example of such observable is the dual of the two-point function of a single trace operator.} We postpone the study of these to future works and only make few comments here:
\item{1.} The T-dual of a closed string carrying momentum $\bk$ in the transverse directions is a string that has a branch cut on the worldsheet and its image in spacetime is no-longer closed. The mismatch in the $\bx$ coordinates \background\ is proportional to $\theta\bk$. In \GrossBA\ it was found that in NCYM, any single trace operator carrying momentum $\bk$ develops a Wilson line attached to it of length $\theta\bk$. We expect these to be the dual of the branched ``closed" strings.
\item{2.} The blindness of NCYM observables to the non-commutativity scale holds only at the planar level. Similarly, the duality between gluon scattering amplitudes and polygon Wilson loops hold only for the planar contribution. However, our identification of the NCAdS dual to NCYM UV observables (specified near the NCAdS boundary in term of the T-dual variables) may hold beyond the planar limit.
\item{3.} We expect a branched closed string near the NCAdS boundary to arise from a flat space brane picture in the following way: Consider a stack of $n$ coincident D3-branes separated from another stack of $N$ coincident D3-branes in flat space. What observable, in the $\alpha'\to 0$ limit of that configuration, reduces to a closed string in AdS that describes a two point correlation function between single trace operators of momentum $\bk$ and $-\bk$? It is the sum of all open string diagrams with any number of holes on the big stack of $N$ branes and only two holes on the stack of $n$ probe D3-branes. In addition, there are corresponding open string vertex insertions along these two boundaries carrying the momentum $\bk$ and $-\bk$. In the near horizon gravitational dual description, that sum is replaced by a closed string in AdS approaching the boundary at two points (integrated against the Fourier transform coefficients). Now suppose we turn on a constant NS $B$-field on the D3-branes world volume and then take the near horizon limit as in \SeibergVS. The constant $B$-field is a total derivative on the open worldsheet. It dresses the boundary of each hole by the $B$-flux through it. For the holes on the big stack of $N$ D3-branes, it is responsible for changing the dual gravitational background from AdS to NCAdS. For the two hole on the stack of probe $n$ D3-branes, it rotates momentum insertions into ``winding". That ``winding" is proportional to $\theta\bk$. Since the total momentum insertion of each of the two holes is non-zero, so is the ``winding". Now the closed string image near the NCAdS boundary is no longer closed. 

\bigskip \bigskip \bigskip \bigskip \bigskip \bigskip
\centerline{\bf{Acknowledgements}}

We thank O. Aharony, F. Cachazo, J. Gomis, X. Liu, R. Myers and T. Okuda  for
discussions and comments. We thank J. Maldacena and J. McGreevy for comments on the manuscript. This research is supported by the Government of Canada through Industry Canada and by the Province of Ontario through the Ministry of Research \& Innovation.

\bigskip
\listrefs
\end